\documentclass[10pt]{article}

\usepackage{a4wide}
\usepackage{latexsym}
\usepackage{amssymb}
\usepackage{amsmath}
\usepackage[title,titletoc]{appendix}

\begin{document}

\def\d{{\rm d}}
\def\ex{{\rm e}}
\def\im{{\rm i}}
\def\u{{\bf u}}
\def\v{{\bf v}}
\def\x{{\bf x}}
\def\g{{\bf g}}
\def\r{{\bf r}}
\def\A{{\bf A}}
\def\s{{\bf s}}
\def\e{{\bf e}}
\def\calR{{\cal R}}
\def\eR{{\bf e}_{\scriptscriptstyle{R}}}
\def\bareR{\bar{\bf e}_{\scriptscriptstyle{R}}}
\def\tildeeR{\tilde{\bf e}_{\scriptscriptstyle{R}}}
\def\U{{\bf U}}
\def\rv{r_{\rm v}}
\def\X{{\bf X}}
\def\balpha{{\boldsymbol\alpha}}
\def\bxi{{\boldsymbol\xi}}
\def\smalze{{\scriptscriptstyle{(0)}}}
\def\smalun{{\scriptscriptstyle{(1)}}}
\def\smaldu{{\scriptscriptstyle{(2)}}}
\def\smaln{{\scriptscriptstyle{(n)}}}
\def\smalm{{\scriptscriptstyle{(m)}}}
\def\smalnze{{\scriptscriptstyle{(n,0)}}}
\def\smalnun{{\scriptscriptstyle{(n,1)}}}
\def\smalndu{{\scriptscriptstyle{(n,2)}}}
\def\beq{\begin{eqnarray}}
\def\eeq{\end{eqnarray}}

\font\brm=cmr10 at 24truept
\font\bfm=cmbx10 at 15truept
\centerline{\brm Mechanical diffusion in grease ice}
\centerline{\brm stirred by gravity waves}
\vskip 20pt
\centerline{Piero Olla}
\vskip 5pt
\centerline{ISAC-CNR and INFN, Sez. Cagliari, I--09042 Monserrato, Italy}
\vskip 20pt

\begin{abstract}
The possibility of hydrodynamic diffusion in a model of grease ice stirred by the velocity field of
a gravity wave is explored. It is argued that mechanical
interactions among ice crystals can induce disturbances in the fluid velocity---in the
form of interstitial flows---analogous to those leading to diffusion in fixed beds. 
A two-fluid description of the system is introduced, in which the
ice matrix is treated as a deformable porous medium. 
%The microscopic structure of the slurry is modeled with a fiber network.
%The analysis is limited to the large Peclet number limit.
Depending on the range of parameters, the effective diffusivity of the medium can 
exceed the value which would be obtained by only keeping into account particle
dislocation induced by contact interactions.

\end{abstract}

%%%%%%%%%%%%%%%%%%%%%%%%%%%%%%%%%%%%%%%%%%%%%%%%%%%%%%%%%%%%%%%%%%%%%%%%%%

\section{Introduction}

Flows in suspensions are often  characterized by transport phenomena that
originate from the interaction of the particles among themselves and with the flow. 
In dilute suspensions, the particle interactions are
mediated by hydrodynamic forces. In more concentrated suspensions, 
the contribution from direct contact of the particles is dominant.
Effective diffusivities greatly exceeding the values expected from 
molecular effects are often observed, even in the absence of turbulence 
\cite{rusconi08,metzger13}. 

A medium of particular interest in the geophysical context
is grease ice, that is the dense slurry of ice crystals often 
observed at the water surface in polar seas during periods of strong wind and freezing 
temperatures \cite{martin81}. Consolidation of the slurry to pack ice rests on efficient
removal of the heat and salt produced in freezing. Due to the high viscosity of the medium,
turbulence is severely hindered. Mechanical diffusion
induced by stirring of the ice by the wind and by gravity waves, on the contrary,
may play an important role. 

A possible mechanism is shear-induced diffusion \cite{eckstein77,leighton87,sierou04}. 
It is possible to make some estimate in the two cases
diffusion is generated by wind stirring and by gravity waves.

The wind will generate in the ice layer a shear of strength $a_s\approx (u_*)^2/\nu_{grease}$,
where $u_*$ is the friction velocity and $\nu_{grease}$ is the effective viscosity of grease ice.
In a concentrated suspension, such as grease ice, particle diffusion will arise from collisions, 
rather than from hydrodynamic interaction between particles. 
Collisions will occur between ice crystals travelling along flow lines separated by
a typical crystal size $L$, corresponding to a collision cross section 
$\sigma\approx L^2$, and to a typical relative velocity in collision $u_L\approx La_s$.
In concentrated conditions, the number density of the crystals is $n\approx L^{-3}$ and
the particle displacement in a collision is $\approx L$. Hence,
the collision rate will be $a_{coll}\approx n\sigma_{coll}u_L\approx a_s$,
and the wind contribution to the diffusivity $\kappa_{wind}\approx a_sL^2$. 
Taking a wind speed of about
10 m/s, corresponding to a friction velocity $u_*\approx 0.01$ m/s \cite{bauer83},
and taking for the grease ice, 
$L\approx 1$ mm and $\nu_{grease}\approx 0.01\ {\rm m^2/s}$ \cite{martin81}, would give
\beq
\kappa^{wind}\approx (u_*L)^2/\nu_{grease}\approx 10^{-8}\ {\rm m^2/s}.
\label{kappa_wind}
\eeq

The contribution to diffusion from gravity waves can be estimated in a similar way.
A crystal in the field of a gravity wave will be able to ``prod'' a nearby crystal only during the 
compressible phase of the wave at the crystal location. The resulting displacement will be
$S\approx u_L/\omega\approx La_s/\omega$, where $\omega$ is the wave frequency,
$a_s\approx\epsilon\omega$ is the strain intensity,  $\epsilon=kU/\omega$ is
the dimensionless wave amplitude, and $k$ and $U$ are the wavenumber and 
the characteristic fluid velocity of the wave. Hence $S\approx\epsilon L$.
In concentrated conditions, a crystal will make collision with a neighbour about once
in a period, and if the degree of irreversibility in collision is sufficient \cite{corte08},
a contribution to the diffusivity, $\kappa^{wave}\approx S^2\omega$, will be generated.
Taking typical values for ocean
waves, $\omega\approx 1\ {\rm s^{-1}}$ and $\epsilon\approx 0.1$, would give
\beq
\kappa^{wave}\approx (\epsilon L)^2\omega\approx 10^{-8}\ {\rm m^2/s}.
\label{kappa_wave}
\eeq
If molecular diffusion is small,
tracers will move with the liquid around the particles, and have the same diffusion 
properties of the particles, described by Eqs. (\ref{kappa_wind}) and (\ref{kappa_wave}).
The diffusivities in Eqs. (\ref{kappa_wind}) and (\ref{kappa_wave}) can then be compared with 
the heat and salinity diffusivities in salt water: 
$\kappa_T\simeq 1.4\times 10^{-7}\ {\rm m^2/s}$ and 
$\kappa_S\simeq 7.4\times 10^{-10}\ {\rm m^2/s}$. The mechanical
contribution to diffusion appears to be relevant only for salinity.

The estimates in Eqs. (\ref{kappa_wind}) and (\ref{kappa_wave}) disregard the possible
impact of relative motions between phases, which may lead to diffusion phenomena analogous
to those generated in flows in fixed particle beds \cite{koch85,koch86}.
Relative motion of the phases is going to be generated in concentrated suspensions
by the interplay between non-hydrodynamic and hydrodynamic (drag) forces
on the particles. In creeping flow conditions, exact balance between mechanical
and drag forces must exist, which corresponds,  macroscopically, to balance
between the drag forces by the interstitial flow and the resistance to deformation in 
the solid matrix.
In diluted suspensions, no mechanical interparticle interactions obviously exist, and 
balance of solid and fluid stresses occurs at the particle scale, that is also 
the scale of the fluid velocity disturbances.

In grease ice, non-hydrodynatic stresses are generated dynamically by the 
continuous formation and breaking of bonds between ice crystals induced by the flow, and
allow the medium to persist in a fluid state.

Note that in order for a crossflow to be present, 
a pressure gradient must exist in the liquid phase. Such pressure gradients are absent
in the homogeneous shear layer generated by the wind stress. They
are generated dynamically, instead, in the field of a gravity wave. 
%This motivates the present study of mechanical diffusion in the field of a gravity wave.
The question remains whether the contribution to diffusion is significant or not.
Answering this question is the goal of the present paper.

To study the problem, a two-phase approach is adopted.  No attempt is made to determine 
the macroscopic stresses of the network from microscopic behaviors, rather, 
an isotropic viscous response to deformation is assumed, with viscosity extrapolated 
from that of real grease ice. Knowing the relative velocity of the phases allows
to obtain diffusivity estimates based on the characteristic size of
the particles in suspension. Not more than an order of magnitude estimate is provided: 
grease ice is a complicated mixture of mm and sub-mm crystals in the form of platelets, 
needles and other shapes, with a solid volume fraction 
$\bar C\approx 0.1$ \cite{bauer83,smedsrud06}.
Analytical treatment of the problem is therefore very difficult.
More precise results on diffusion in the field of a gravity wave are obtained in the Appendix, 
in the idealized case of a dilute fiber network.

The analysis predicts 
significant diffusivity enhancement with respect to the estimates in 
Eqs. (\ref{kappa_wind}) and (\ref{kappa_wave}), with an amplification factor 
proportional to the ratio between
the distance traveled by a tracer in the solid matrix and the crystal scale.
The key condition for amplification is that the effective viscosity of the slurry is large
compared to that of the interstitial liquid. 
The diffusivity dependence on the solid volume fraction is weaker. 

The paper is organized as follows. In Sec. 2, the macroscopic equations describing
the two-phase system are derived. In Sec. 3, the equations are applied to the 
calculation of the relative motion between phases.
In Sec. 4, the contribution to diffusivity is evaluated.
Section 5 is devoted to conclusions. Analytical results for diffusion in a dilute random 
fiber bed are presented in the Appendix.

\section{Two-phase model}
The dynamics of gravity waves in grease ice covered ocean is studied by modelling the ocean 
as an infinitely deep uniform slurry. As the thickness of grease ice in the ocean, and the depth 
of the region where wave dissipation takes place, are usually comparable 
\cite{longuet53,desanti17}, the model is going to be rough, but not completely inaccurate. 

A macroscopic two-phase description is adopted, with the solid and liquid 
phases obeying separate mass and momentum conservation equations.
No hypothesis is made for the moment on the microscopic structure of the medium,
except for the fact that non-hydrodynamic interactions
dominate the stress in the ice matrix.
%In grease ice, non-hydrodynamic interactions are generated by the continuous formation
%and breaking of bonds between ice crystals induced by the flow. Such interactions
%have predominantly dynamical origin, from the continuous collisions between ice crystals,
%and allow the medium to persist in a fluid state.

To fix the ideas, take the $x$ axis along the
direction of propagation of the wave, and the $z$ axis pointing upwards, with 
$z=0$ at the surface. 

The momentum and continuity equations obeyed by the two phases read:
\beq
&&(1-\bar C)[\rho_l\frac{\partial\U_l}{\partial t}+\nabla P
-\mu\nabla^2\U_l]=\Gamma(\U_s-\U_l)+(1-C)\rho_l\g,
\label{U_l}
\\
&&\bar C[\rho_s\frac{\partial\U_s}{\partial t}+\nabla P-\mu\nabla^2\U_l]
=-\nabla P_s+\mu_s^{\scriptscriptstyle{B}}\nabla\nabla\cdot\U_s+\mu_s\nabla^2\U_s-\Gamma(\U_s-\U_l)+C\rho_s\g,
\label{U_s}
\\
&&\frac{\partial\tilde C}{\partial t}+\bar C\nabla\cdot\U_s=0,
\label{continuity}
\\
&&\nabla\cdot[(1-\bar C)\U_l+\bar C\U_s]=0,
\label{incompressibility}
\eeq
where linearity has been imposed, tilde and overbar indicate fluctuating and mean 
quantities, $g\simeq 9.8\ {\rm m/s^2}$, $\Gamma$ and $C$ are the gravitational
acceleration, the drag coefficient and the solid volume fraction, and
$\rho_n$, $\U_n$ and $\mu_n$ are the mass density, the velocity
and the dynamic viscosity of the liquid ($n=l$) and solid ($n=s$) phase.
%(see \cite{morris09} and references therein for an account of the use of 
%two-phase models in the study of suspension dynamics).
Similar equations have been used in \cite{spiegelman93} to model the flow in 
deformable porous media.

The terms 
$\nabla P_s$, $\mu_s^{\scriptscriptstyle{B}}\nabla\nabla\cdot\U_s$ and $\mu_s\nabla^2\U_s$ 
in Eq. (\ref{U_s})
give the
non-hydrodynamic part of the pressure and viscous forces. 
All three contributions arise from coarse graining of the particle mechanical 
interactions. Note that when
these terms are zero, and the particles are neutrally buoyant, 
Eqs. (\ref{U_l}) and (\ref{U_s}) have solution $\U_l=\U_s$.
The mechanical pressure $P_s$ and
the bulk viscosity $\mu_s^{\scriptscriptstyle{B}}$ account for the resistance to compression of the solid matrix.
An isotropic expression is utilized for the bulk viscosity
term even though such normal stresses are typically anisotropic in interacting
particle suspensions \cite{morris99}. As the flow of the individual phases
is almost incompressible (see below), the expected error is small. 

The terms $\nabla P$ and $\mu\nabla^2\U_l$ in Eqs. (\ref{U_l}) and (\ref{U_s}) 
give the hydrodynamic part of the pressure and
viscous forces, with $\mu\nabla^2\U_l$ receiving contribution from the stress both in the
liquid and in the particles (the Einstein correction in a dilute suspension). 
The hydrodynamic pressure $P$
is determined by the incompressibility condition Eq. 
(\ref{incompressibility}). The mechanical pressure  is taken to obey a
constitutive relation in the form $P_s=P_s(C;\nabla\U_s)$, and to rapidly vanish for
$C>\bar C$, to allow hydrostatic equilibrium at $C\approx\bar C$. 

Hydrostatic equilibrium for both phases is obtained by
setting $\U_{l,s}=0$ in Eqs. (\ref{U_l}) and (\ref{U_s}). 
The result is, assuming slow variation of $\bar C$ throughout the domain,
\beq
\bar P_s(z)=\bar P_s(0)+\bar Cg(\rho_l-\rho_s)z
\quad{\rm and}\quad
\bar P(z)=\bar P(0)-g\rho_lz.
\label{buoyancy}
\eeq
(Note that in the absence of a mechanical pressure, 
$\bar P(z)=\bar P(0)-g[\rho_l+\bar C(\rho_s-\rho_l)]z$,
buoyancy forces on the two phases,
$-\bar f_l=\bar f_s=-\bar C(\partial_z\bar P+\rho_sg)=g(1-\bar C)\bar C(\rho_l-\rho_s)$,
would be present, and would need to be balanced by the drag forces generated by 
settling of the particles to the surface.)
%\beq
%\bar P(z)=-\bar P_s(z)-g[\rho_l+\bar C(\rho_s-\rho_l)]z.
%\label{hydrostatic}
%\eeq
%The force densities on the individual phases are
%\beq
%&&\bar f_s=-\partial_z\bar P_s-\bar C\rho_sg=g(1-\bar C)[\partial_z\bar P^M-\bar C(\rho_l-\rho_s)],
%\\
%\nonumber
%&&\bar f_l=-\partial_z\bar P_l-(1-\bar C)\rho_lg=
%-g(1-\bar C)[\partial_z\bar P^M-\bar C(\rho_l-\rho_s)].
%\nonumber
%\eeq

Subtraction of the hydrostatic component accounted for by Eq. (\ref{buoyancy}), 
from Eqs. (\ref{U_l}) and (\ref{U_s}), gives
\beq
&&(1-\bar C)[\rho_l\frac{\partial\U_l}{\partial t}+\nabla \tilde P-\mu\nabla^2\U_l]=
\Gamma(\U_s-\U_l), \label{U_l1}
\\
&&\bar C[\rho_s\frac{\partial\U_s}{\partial t}+\nabla\tilde P-\mu\nabla^2\U_l]
=-\beta\nabla\tilde C+\mu_s^{\scriptscriptstyle{B}}\nabla\nabla\cdot\U_s+\mu_s\nabla^2\U_s
\nonumber
\\
&&\qquad\qquad\qquad\qquad\qquad\qquad\quad\ 
-\Gamma(\U_s-\U_l)+\tilde C(\rho_s-\rho_l)\g,
\label{U_s1}
\eeq
where 
\beq
\beta=(\partial P_s/\partial C)_{C=\bar C}
\label{compressibility}
\eeq
is the compressibility of the solid phase.

In the case of monocromatic waves, one can write for the fluctuating quantities:
\beq
\tilde C(\r,t)=\tilde C(z)\ex^{\im(kx-\omega t)},
\qquad
\tilde C(z)=\tilde c_p\ex^{kz}+\sum_n\tilde c_n\ex^{\alpha_nz},\ldots,
\label{tilde C}
\eeq
where subscripts $p$ and $n$ identify potential and non-potential components.
The exponents $\alpha_n$ give the depth of the boundary layers induced by
the two-phase dynamics at the water surface.

The velocities can be expressed in terms of velocity potentials \cite{lamb}:
\beq
\U_s=-\nabla\Phi-\nabla_\perp A,
\qquad
\U_l=\U_s-\nabla\hat\Phi-\nabla_\perp\hat A,
\label{potentials}
\eeq
where $\nabla_\perp=(\partial_z,-\partial_x)$. The fields $A$ and $\tilde A$ are
chosen to contribute only to the vortical part of $\U_{l,s}$, and therefore
do not have potential components $A_p=a_p\ex^{kz}$ and $\hat A_p=\hat a_p\ex^{kz}$.
Such components could be eliminated anyway by a gauge transformation $\Phi\to\Phi-\im A_p$,
$\hat\Phi\to\hat\Phi-\im\hat A_p$.

Substitution of Eq. (\ref{potentials}) into 
Eqs. (\ref{continuity}) and (\ref{incompressibility}) gives
\beq
\tilde c_p= 0,
\qquad
\tilde c_n= -\frac{(\alpha_n^2-k^2)\bar C}{\im\omega}\phi_n
\quad{\rm and}\quad
\phi_n= -(1-\bar C)\hat\phi_n.
\label{pippo0}
\eeq
Equations for the non-potential components $\tilde p_n$, $\hat\phi_n$, $a_n$ and $\hat a_n$ are
obtained by taking $\nabla_\perp$ and $\nabla$ of Eqs. (\ref{U_l1}) and (\ref{U_s1}), 
and then using Eq. (\ref{pippo0}):
\beq
&&(1-\bar C)[\im\omega\rho_l+\mu(\alpha_n^2-k^2)](a_n+\hat a_n)
-\Gamma\hat a_n=0,
%-\frac{g\rho_l\bar Ck}{\omega}(1-\bar C)\hat\phi_n=0,
\label{pippa1}
\\
&&\bar C[\im\omega\rho_sa_n+\mu(a_n+\hat a_n)]
+\mu_s(\alpha_n^2-k^2)a_n
+\Gamma\hat a_n-\frac{g(\rho_l-\rho_s)\bar Ck}{\omega}(1-\bar C)\hat\phi_n=0,
\label{pippa2}
\\
&&
\{[\im\omega\rho_l+\mu(\alpha_n^2-k^2)
%-\frac{\rho_lg\bar C\alpha_n}{\im\omega}
]\bar C(1-\bar C)
-\Gamma\}\hat\phi_n+(1-\bar C)p_n=0,
\label{pippa3}
\\
&&\{[-\im\omega\rho_s\bar C
-(\mu_s^{\scriptscriptstyle{B}}+\mu_s-\bar C\mu-\frac{\beta\bar C}{\im\omega})(\alpha_n^2-k^2)
\nonumber
\\
&&\qquad\quad-\frac{g(\rho_l-\rho_s)\bar C\alpha_n}{\im\omega}](1-\bar C)+\Gamma \}\hat\phi_n
+\bar C\tilde p_n=0.
\label{pippa4}
\eeq

The potential terms are obtained by integrating Eqs. (\ref{U_l}) and (\ref{U_s})
over $z$, and then exploiting Eqs.  (\ref{buoyancy}) and (\ref{pippo0}-\ref{pippa4}). The result
is
\beq
&&
(1-\bar C)[\im\omega\rho_l(\phi_p+\hat\phi_p)
+p_p+\rho_lg\eta_l]
-\Gamma\hat\phi_p=0,
\label{pippa5}
\\
&&
\bar C[\im\omega\rho_s\phi_p+p_p+\rho_sg\eta_s]+\Gamma\hat\phi_p=0,
\label{pippa6}
\eeq
where the vertical displacement at the surface, $\eta^{s,l}$, obeys
\beq
-\im\omega\eta^{s,l}=U_z^{s,l}|_{z=0}.
\label{pippa7}
\eeq
For weakly damped waves (which is the case in most situations of interest \cite{newyear99}),
$k\simeq k_\omega=\omega^2/g$, and the dominant contribution to $\U_l\simeq\U_s$ is produced by 
$\phi_p$. The
velocity scale of the waves is therefore $U=k_\omega|\phi_p|$.

\section{Determining the crossflow}
It is convenient to shift to dimensionless variables. The various 
quantities appearing in Eqs. 
(\ref{U_l}-\ref{incompressibility}) and below are rescaled as follows:
$\omega t\to t$, $k/k_\omega\to k$, $\mu/\mu_s\to\mu$, 
$\sqrt{\mu_s/(\omega\rho_l)}\alpha\to\alpha$, 
$\beta\bar C/(\omega\mu_s)\to\beta$,
$\Gamma/(\omega\rho_l)\to \Gamma$, $pk_\omega/(\rho_lg)\to p$,
$Uk_\omega/\omega\to U$.
The velocity scale of the waves becomes $U=\epsilon$.

The following
dimensionless parameters  are introduced:
\beq
\gamma=1+\frac{\mu_s^{\scriptscriptstyle{B}}}{\mu_s},
\qquad
\rho=\frac{\rho_s}{\rho_l},
\quad{\rm and}\quad
\nu_g=
\frac{k_\omega^{3/2}\mu_s}{\rho_lg^{1/2}},
\label{dimensionless}
\eeq
and the following smallness conditions are imposed
\beq
\mu,\ \nu_g,\ \bar C \ll 1.
\label{small}
\eeq
The stresses in the solid matrix are assumed to be of comparable magnitude; therefore
$\gamma,\beta=O(1)$. 

The conditions in Eq. (\ref{small}) correspond, in the order, to the requirements:
\begin{itemize}
\item
high viscosity of the medium compared to that of the liquid phase; 
\item
waves that are only weakly damped \cite{keller98,desanti17}, in such a way 
that $|k-1|\ll 1$ and all velocity potentials
are small except $\phi_p$;
\item
diluteness of the medium.
\end{itemize}
The first two conditions are easily satisfied by grease ice in the ocean. 
Assuming that all of the viscosity
renormalization comes from mechanical interactions, in such a way that $\mu\approx\mu_l=
\rho_l\nu_l$, $\nu_l\simeq 1.5\times 10^{-6}\ {\rm m^2/s}$, and taking for the viscosity
in the solid matrix
$\nu_g\approx\nu_{grease}\approx 0.01\ {\rm m^2/s}$ \cite{martin81}, would give
$\mu\approx 10^{-4}$. In similar fashion, 
considering wavelengths in the range of $10-100$ m, would give for the dimensionless
viscosity,
$1.5\times 10^{-3}>\nu_g>5\times 10^{-5}$. 

The last condition $\bar C\ll 1$ is only 
marginally satisfied, and is adopted for simplicity. An upper bound for the grease ice 
concentration is fixed by the transition from grease ice to compact 
ice, which takes place at $\bar C\approx 0.3$ \cite{maus12}.

Working with dimensionless variables and expanding to lowest order in the small quantities,
Eqs.  (\ref{pippa1}-\ref{pippa4}) take the form
\beq
&&\im a_n-\Gamma\hat a_n\simeq 0,
\label{peppe1}
\\
&&[\im\rho\bar C+\alpha_n^2]a_n
+\Gamma\hat a_n-(1-\rho)\bar C\hat\phi_n\simeq 0,
\label{peppe2}
\\
&&\Gamma\hat\phi_n-p_n\simeq 0,
\label{peppe3}
\\
&&[\im(1-\rho)\nu_g^{-1/2}\bar C\alpha_n-(\gamma+\im\beta)\alpha^2_n
+\Gamma]\hat\phi_n+\bar Cp_n\simeq 0
\label{peppe4}
\eeq
From Eqs. (\ref{peppe1}-\ref{peppe4}) an 
incompressible mode for the individual phases is identified:
\footnote{Note that in rescaled variables a factor $\nu^{-1/2}_g$ is extracted
out of $\alpha_n$, in such a way that the $z$-derivative of
non-potential fields $Q_n$ is $\partial_zQ_n=\nu_g^{-1/2}\alpha_nQ_n$.} 
\beq
\alpha_i\simeq\sqrt{-\im},
\quad
\hat a_i\simeq\im a_i/\Gamma,
\quad 
\hat\phi_i\simeq p_i\simeq 0.
\label{incompressible}
\eeq
It is easy to recognize in this mode the standard behavior of the vortical component 
of the velocity field of a wave propagating in a viscous medium of  
viscosity $\nu_g$ \cite{lamb,desanti17}.

Equations (\ref{peppe1}-\ref{peppe4}) have an additional compressible mode with eigenmodes
\beq
\alpha_c\simeq\frac{\im(1-\rho)\bar C\pm\sqrt{4(\gamma+\im\beta)\Gamma\nu_g-(1-\rho)^2\bar C^2}}
{2(\gamma+\im\beta)\sqrt{\nu_g}},
\label{alpha_c}
\eeq
where only the root with a positive real part is admissible in an infinitely
deep domain. The root's inverse gives the 
thickness of the compressible boundary
layer. From Eqs. (\ref{incompressible}) and (\ref{alpha_c}), 
$|\alpha_c/\alpha_i|\sim\max(\Gamma^{1/2},\bar C\nu_g^{-1/2})\gg 1$, which implies that the 
compressible boundary layer is exceedingly thin. 

From Eqs. (\ref{peppe1}-\ref{peppe4}) one
can then write for the compressible component of the velocity potentials and of the pressure:
\beq
p_c\simeq \Gamma\hat\phi_c,
\quad
a_c\simeq \frac{(1-\rho)\bar C}{\alpha_c^2}\hat\phi_c,
\quad
\hat a_c\simeq 
\frac{\im(1-\rho)\bar C}{\alpha_c^2\Gamma}\hat\phi_c.
\label{compressible}
\eeq

To derive a dispersion relation, it is necessary to impose
boundary conditions at the fluid free surface, and exploit Eqs.
(\ref{pippa5}-\ref{pippa7}) to express the amplitude of non-potential modes
in terms of potential ones. There are in total five unknowns:
$k$, $p_p$, $\hat\phi_p$, $\hat\phi_c$ and $a_i$; $\phi_p$ remains arbitrary
because of linearity. Thus, three boundary conditions are required. Two are 
well known from the study of gravity waves in viscous media \cite{lamb}, and are the
requirement that the normal stress
$\tau_{zz}\simeq-\tilde P-\tilde P_s+2\mu_s\partial_zU_{sz}+\mu_s^{\scriptscriptstyle{B}}\nabla\cdot\U_s$
and the tangential stress $\tau_{xz}\simeq\mu_s(\partial_xU_{sz}+\partial_zU_{sx})$
at the water surface are identically zero. A reasonable third boundary condition 
is that the interfaces with the atmosphere of the two phases move together,
$[U_s-U_l]_{z=0}=0$, so that there are no sloshing phenomena at the water surface.
Working with velocity  potentials,
\beq
[U_s-U_l]_{z=0}\simeq\hat\phi_p+ \alpha_c\nu_g^{-1/2}\hat\phi_c=0,
\label{U-U}
\eeq
and therefore,
\beq
\eta_s\simeq\eta_l\simeq -\im k\phi_p-a_i+\im\alpha_c\nu_g^{-1/2}\hat\phi_c.
\label{eta}
\eeq
Substitution of Eqs. (\ref{U-U}) and (\ref{eta}) into the equations for the potential
terms, Eqs. (\ref{pippa5}) and (\ref{pippa6}), gives
\beq
&&
\im(1-k)\phi_p+\tilde p_p-a_i+\Gamma\nu_g^{-1/2}\alpha_c\hat\phi_c=0,
\\
&&\bar C[\im\rho(1-k)\phi_p+\tilde p_p-\rho a_i]-\Gamma\nu_g^{-1/2}\alpha_c\hat\phi_c=0,
\eeq
which allows to determine the amplitude of the compressible mode as a function of potential
modes:
\beq
\hat\phi_c\simeq\frac{(1-\rho)\bar C\nu_g^{1/2}}{\Gamma\alpha_c}\tilde p_p,
\qquad
\tilde p_p\simeq -\im(1-k)\phi_p+a_i.
\label{hat phi_c}
\eeq
The compressible mode is driven by buoyancy.
The system of equations is closed by imposing the conditions on the normal and tangential
stress at the surface:
\beq
&&\tau_{zz}|_{z=0}\simeq
-\tilde p_p+2\im\nu_g^{1/2}\alpha_ia_i+[(\im\beta+\gamma+1)\alpha_c^2+2\im(1-\rho)\nu_g^{1/2}
\bar C\alpha_c^{-1}]\hat\phi_c=0,
\label{tau_zz}
\\
&&\tau_{xz}|_{z=0}\simeq
-2\im\nu_g\phi_p+\im a_i+[(1-\rho)\bar C+2\im \nu_g^{1/2}\alpha_c]\hat\phi_c=0.
\label{tau_xz}
\eeq
Substitution of Eq. (\ref{hat phi_c}) into Eqs. (\ref{tau_zz}) and (\ref{tau_xz}) gives
finally
\beq
k\simeq 1+4\im\nu_g,
\qquad
a_i\simeq 2\nu_g\phi_p,
\qquad
\hat\phi_c\simeq -\frac{2(1-\rho)\bar C\nu_g^{3/2}}{\Gamma\alpha_c}\phi_p.
\label{reldisp}
\eeq
One recovers the dispersion relation and the equation for the boundary layer structure of 
gravity waves in a homogeneous medium of kinematic viscosity 
$\nu_g\simeq\mu_s/\rho_l$ \cite{lamb}.
The suspension inherits the viscosity of the solid phase, as expected.
Similar mechanisms have been invoked in \cite{mills95} to explain the rheology
of concentrated sphere suspensions. 
The compressible mode does not contribute to lowest order to the dynamics, as can 
be verified by back substitution of Eqs. (\ref{reldisp}) and (\ref{alpha_c}) 
into Eqs. (\ref{tau_zz}) and (\ref{tau_xz}). This suggests that use of a boundary condition
on $\eta^{s,l}$ could be avoided by simply setting $\hat\phi_c=0$, thus neglecting
all contributions from the compressible mode. The thin compressible boundary layer at the
surface would be replaced in this way by a finite gap between phases, whose amplitude
can be evaluated from Eqs. (\ref{U-U}) and (\ref{reldisp}):
$\eta_s-\eta_l\simeq 2(1-\rho)\bar C\nu_g\Gamma^{-1}\phi_p$. 

It is now possible to determine the relative motion of the phases. Inspection of Eqs.
(\ref{incompressible}), (\ref{compressible}), (\ref{U-U}) and (\ref{reldisp}) allows
to conclude that the dominant contribution is produced by the vortical incompressible
mode $\hat a_i$. This leads to the final result
\beq
\U_l-\U_s\simeq -(\im\sqrt{-\im},\nu_g^{1/2})\frac{2\nu_g^{1/2}}{\Gamma}\phi_p.
\label{U_l-U_s}
\eeq
The relative motion between phases is predominantly along the horizontal, and is confined
in a region of thickness $\lambda_\alpha\approx\nu_g^{1/2}$ near the surface.
The incompressible mode dominates the dynamics both with regard to the corrections to wave
propagation (through $a_i$) and to the interaction between solid and liquid phase 
(through $\hat a_i$).

\section{Diffusivity estimate}
Knowing the crossflow $\U_l-\U_s$ allows to estimate the medium diffusivity
in terms of the characteristic scale $L$ of the crystals
\beq
\kappa^{wave}\approx |\U_l-\U_s|L.
\label{kappa}
\eeq 
A difficulty arises however because of the oscillatory character of $\U_l-\U_s$.
In creeping flow conditions, reversibility of Stokes flows would lead, in the
absence of mechanical interactions of the crystals, to pure oscillatory motion 
of the crystals and of the interstitial liquid; tracer trajectories would
be closed and no diffusion would be possible. In order to have diffusion,
the effect of inertia and of mechanical interactions must be taken into account,
with chaos in the tracer trajectories possibly helping in the process. In this
regards, arguments in \cite{lester13} suggest that tracer excursions in the solid 
matrix of the order of just a few crystal spacings could be enough to produce 
trajectory decorrelation. If a condition of large tracer excursion compared to the 
crystal spacing is satisfied, $\lambda\approx\omega^{-1}|\U_l-\U_s|\gg L$, 
diffusive behaviours are therefore expected, and
Eq. (\ref{kappa}) applies.

Considering waves of normalized amplitude $\epsilon$,
substitution
of Eq. (\ref{U_l-U_s}) into Eq. (\ref{kappa}) gives, after making $L$ dimensionless
by rescaling $k_\omega L\to L$, 
\beq
\kappa^{wave}\approx \frac{\nu_g^{1/2}L\epsilon}{\Gamma}.
\label{kappa_zz}
\eeq
The large excursion condition $\lambda\gg L$ 
becomes, using again Eq. (\ref{U_l-U_s}), 
\beq
\nu_g^{1/2}\epsilon\gg\Gamma.
\label{cond0}
\eeq
The drag coefficient in Eqs. (\ref{kappa_zz}) and (\ref{cond0}) can be estimated
as the product $\Gamma\approx nF$ of the drag force on an individual particle 
$F\approx \mu_lL$ and the particle density $n\approx L^{-3}$:
\beq
\Gamma\approx \frac{\mu\nu_g}{L^2}.
\label{Gamma}
\eeq
The condition $\lambda\gg L$, Eq. (\ref{cond0}),  becomes therefore
\beq
\calR=\frac{\lambda}{L}\approx\frac{\epsilon L}{\mu\nu_g^{1/2}}\gg 1,
\label{calR}
\eeq
and the diffusivity in Eq. (\ref{kappa_zz}), back to dimensional units, takes the form
\beq
\kappa^{wave}\approx \calR L^2\omega.
\qquad
\label{kappa_zz1}
\eeq
The amplification factor with respect to the shear-induced part of the diffusion coefficient
in Eq. (\ref{kappa_wave}) is $\calR/\epsilon^2$. 
%The dependence on $\calR$ of the 
%amplification factor is remarkable, and contains information on the two-phase flow in 
%the field of a gravity wave which cannot be obtained from dimensional calculus.
The diffusivity is independent of the volume fraction $\bar C$. A more precise
calculation carried out in the Appendix, in the case of a dilute fiber suspension,
confirms the result within logarithmic corrections.

The estimate in Eq. (\ref{kappa_zz1}) can be applied to the case of gravity waves in grease
ice. In the case of 100 m long waves, 1 mm ice crystals, and an ice viscosity 
$\nu_g\approx\nu_{grease}\approx 0.01\ {\rm m^2/s}$,  corresponding in dimensionless variables to
$\nu_g\approx 5\times 10^{-5}$ and $\Gamma\approx 2$, Eqs. (\ref{calR}) and (\ref{kappa_zz1})
would give
\beq
\calR\approx 15,
\quad{\rm and}\quad
\kappa^{wave}\approx 10^{-5}\ {\rm m^2/s}.
\label{calR1}
\eeq
Using the same ice parameters in the case of 10 m waves, corresponding in dimensionless variables 
to $\nu_g\approx 1.5\times 10^{-3}$ and $\Gamma\approx 0.6$, 
would give instead $\calR\approx 16$ and 
$\kappa^{wave}\approx 4\times 10^{-5}\ {\rm m^2/s}$.

Note that the diffusivity increment described in Eq. (\ref{calR1}) is localized in the viscous
boundary layer of thickness $\lambda_\alpha=\sqrt{\nu_g/\omega}$ near the water surface, which,
as already remarked, coincides in most situations of interest with the thickness of the
grease ice layer.

\section{Concluding remarks}
The analysis carried out in the present paper predicts a mechanical contribution to heat and
mass diffusion in grease ice stirred by a gravity wave, that exceeds by several orders of magnitude
the contribution by shear-induced diffusion. The 
corresponding increment over the molecular diffusivity is $O(10^5)$ in the case of salinity,
$O(10^2)$ in the case of heat. Analysis in the Appendix suggests that
the contribution of prefactors not included in the analysis may modify the estimates by
$O(10^{-2})$ factors, thus it is not possible to conclude that heat diffusion is
strongly
affected. Less likely that the prefactor contribution is so huge to lead to similar conclusions
in the case of salinity.  The predicted enhancement of haline diffusion
may lead to a shift in the onset of haline convection during ice formation \cite{wettlaufer97}.

The mechanism of diffusion described in this paper is not confined to suspension
stirring by gravity waves. The mechanism is expected to be active in all situations in which
pressure gradients in the flow and non-hydrodynamic interactions of the particles are
important. 

A question which arises in the case of periodic flows
is how much does Stokes flow reversibility in creeping flow conditions
affect the development of diffusion.
A similar question has attracted recently great attention in the case of shear-induced particle
diffusion in periodic shear flows \cite{pine05}. In that case, irreversibility
is restored by particle collisions, provided the shear is above a concentration dependent threshold
\cite{pham15}. In the present case, it is reasonable to expect that a combination of
chaos in the tracer trajectories, particle inertia, and dislocations induced by the flow, all
play a role in producing the necessary irreversibility for tracer diffusion. The observation
is strengthened by the predicted dependence of the diffusivity on the tracer
excursion in a wave period in the solid bed compared to the crystal separation.
Whether or not some threshold for irreversibility exists, analogous to the one in the case
of shear-induced diffusion, remains to be ascertained.

\vskip 10pt
\noindent{\bf Aknowledgements}: This research was supported by FP7 EU project ICE-ARC
(Grant agreement No. 603887).

\begin{appendices}
\section{Diffusion in a dilute fiber bed}
Transverse diffusion in flows in fixed particle beds is possible only if the particles are 
non-spherical, or if $O(\bar C^2)$ effects are taken into account.
The only analytical results available in the first case are those for flows
in dilute fiber beds \cite{koch86}. The case of a random fiber network is considered here, with
contact among fibers providing the necessary source of non-hydrodynamic interactions.
The length $L$ now indicates the distance along a fiber between different points of contact
with other fibers. The volume fraction is expressed in terms of the fiber radius $r$ 
as $\bar C\approx (r/L)^2$. 

The drag on the liquid phase is obtained, in the dilute limit,
from that of an individual fiber, by using slender body theory \cite{batchelor70,spielman68}.
In the case of random fiber orientation,
\beq
\Gamma\approx\frac{20\pi\mu\nu_g}{3|\ln\bar C|L^2}.
\label{Gamma1}
\eeq
In order for some form of mechanical diffusion to be present, it is necessary that the
displacement $\lambda$ of a tracer relative to the network in a wave period greatly exceeds the
correlation length of the velocity fluctuations in the medium. This correlation length can be
identified with the Brinkman length \cite{spielman68}, 
\beq
\lambda_B=\sqrt{\frac{\mu\nu_g}{\Gamma}}\approx\sqrt{\frac{3|\ln\bar C|}{20\pi}}L,
\label{lambda_B}
\eeq
that is the length above which the velocity perturbation around a
fiber is exponentially damped by the image field of the other fibers.
In the case of a dilute network, $\lambda_B> L$.
Redefining $\calR=\lambda/\lambda_B$, and exploiting Eqs. (\ref{U_l-U_s}) and (\ref{Gamma1}) 
with $\phi_p=O(\epsilon)$, gives
\beq
\calR\approx
\sqrt{\frac{3|\ln\bar C|}{5\pi}}\frac{\epsilon L}{\mu\nu_g^{1/2}}\gg 1.
\label{cond1}
\eeq
The transverse diffusivity of the medium is given in the large Peclet number by
Eq. (31b) in \cite{koch86}:
\beq
\kappa^{wave}_{zz}\approx\kappa^{wave}_{yy}
\approx\frac{9\pi^3}{6400}\frac{a^2}{\lambda_B\bar C}|\U_l-\U_s|.
\label{Koch & Brady}
\eeq
Exploiting Eqs. (\ref{U_l-U_s}) and (\ref{lambda_B}) gives 
\beq
\kappa^{wave}_{zz}\approx \kappa^{wave}_{yy}\approx\frac{9\pi^3}{6400}\calR L^2\omega,
\label{D}
\eeq
corresponding to a diffusivity reduction with respect to the prediction 
in Eq. (\ref{kappa_zz1}), by a factor $\approx 0.02\,|\ln \bar C|^{1/2}$.
\end{appendices}

\end{document}